\documentclass[aps,twocolumn,amssymb,superscriptaddress]{revtex4}

\usepackage{epsfig, color}
\usepackage{amsthm, amsmath, amsfonts, ae, psfrag, bbold}

\begin{document}

\title{Theory of ice premelting in porous media}

\author{Hendrik Hansen-Goos}
 \email{hendrik.hansen-goos@yale.edu}

\affiliation{Department of Geology and Geophysics, Yale University, New Haven, CT 06520, USA
}

\author{J.~S.~Wettlaufer}

\affiliation{Department of Geology and Geophysics, Yale University, New Haven, CT 06520, USA
}

\affiliation{Department of Physics, Yale University, New Haven, CT 06520, USA}

\affiliation{Program in Applied Mathematics, Yale University, New Haven, CT 06520, USA}

\date{\today}   

\begin{abstract}

Premelting describes the confluence of phenomena that are responsible for the stable existence of the liquid phase of matter in the solid region of its bulk phase diagram.  Here we develop a theoretical description of the premelting of water ice  contained in a porous matrix, made of a material with a melting temperature substantially larger than ice itself, to predict the amount of liquid water in the matrix at temperatures below its bulk freezing point. Our theory combines the interfacial premelting of ice in contact with the matrix, grain boundary melting in the ice, and impurity and curvature induced premelting, the latter occurring in regions which force the ice-liquid interface into a high curvature configuration. These regions are typically found at points where the matrix surface is concave, along contact lines of a grain boundary with the matrix, and in liquid veins. Both interfacial premelting and curvature induced premelting depend on the concentration of impurities in the liquid, which, due to the small segregation coefficient of impurities in ice are treated as homogeneously distributed in the premelted liquid.  Our principal result is an equation for the fraction of liquid in the porous medium as a function of the undercooling, which embodies the combined effects of interfacial premelting, curvature induced premelting, and impurities. The result is analyzed in detail and applied to a range of experimentally relevant settings. 

\end{abstract}


\maketitle

\section{Introduction}

Due to advances in our understanding of the premelting of ice and other materials over the last fifteen years \cite{DRW06}, and an increasing appreciation of the influence and extent of this phenomenon in technological, terrestrial and extraterrestrial settings, it is prudent to revisit the detailed thermodynamic basis of its manifestation in porous media where water is invariably found in adsorbed layers or in bulk.  In 1992 Cahn, Dash, and Fu (CDF) \cite{CaDaFu92} developed the original premelting based theory for the liquid fraction $f_l$ of water in a porous medium at sub-freezing temperatures. Given experimental and theoretical advances in the interim, we extend their work in a number of respects. CDF considered a matrix of monodisperse spheres arranged either in a simple cubic packing or in a cubic close packing, regarded as lower and upper bounds for the experimental systems. We design our theory for the {\em random} close packing of monodisperse spheres, which may be a more realistic description for most applications, e.g.\ frozen soils, where even a small degree of polydispersity prevents a periodic packing of the matrix particles. The assumed periodic packing of the matrix particles leads CDF to take the grain size of the interstitial ice to be on the order of the size of the matrix lattice unit cell. Given that the degree of polycrystallinity of the interstitial ice has a complicated dependence on the freezing rate \cite[see e.g., Ref.][and references therein]{PMW10} and the sample annealing history, the ice grain size may differ substantially from the matrix particle size. Whence, polycrystallinity of the interstitial ice enters into our theory as a parameter, namely the density of grain boundary surface relative to the interstitial volume.  Calculations since CDF showed that the dispersion force contribution to interfacial melting could be repulsive or attractive \cite{WiWeElSch95, DRW06} and thus favor or suppress melting in a manner and strength that depends on the effective dielectric response of the ice/water/matrix interface.  Moreover, it has been understood that the great inter-experimental variability in both the interfacial and surface melting of ice was likely due to the sensitivity of the phenomena to impurities which have a rich and complex influence on the melting behavior \cite{We99}.
Thus, our theory allows for both attractive and repulsive dispersion forces of arbitrary strength (depending on the matrix material) and includes the effect of charges at the interface, which give rise to an electrostatic force.  Finally, most importantly, our work includes the effect of charged impurities in the premelted liquid on both interfacial and curvature induced premelting. 

The article is structured as follows. In Section~\ref{sec_theory} we lay out the tenets and details of our theory, discussing interfacial premelting and curvature induced premelting in the presence of charged impurities and combine them in order to describe premelting in a porous medium. The final result of the theoretical treatment is an implicit functional equation for the liquid fraction $f_l$, Eq.~\eqref{eq_fltot}. The first part of Section~\ref{sec_analysis} is dedicated to the analysis of interfacial premelting as described by Eq.~\eqref{eq_explinterf}, and in the second part  Eq.~\eqref{eq_fltot} for the liquid fraction $f_l$ is analyzed by extracting exact solutions for certain limiting cases and numerical solution in the general case. The porous media which we consider consist of fused quartz, gold, and silicon. We summarize in Section~\ref{sec_sumconc}.

\section{Theory}

\label{sec_theory}

\subsection{Interfacial melting and grain boundary melting: Planar interfaces}

\label{sec_mechLayer}

In order to calculate the equilibrium thickness $d$ of a premelted layer for a given undercooling $\Delta T = T_m - T$, where $T_m$ is the bulk melting temperature of ice and $T$ is the temperature of the system, we consider a three layer planar system. Layer $I$ is given by the substrate, i.e., in the case of interfacial melting, the layer consists of the material which forms the porous medium, or of ice when grain boundary melting is considered. Layer $I\!I$ is the premelted quasi-liquid layer of thickness $d$ and finally layer $I\!I\!I$ consists of ice.   In what follows we consider the complete free energy of the system and then focus on the individual complications and contributions in turn to show how known asymptotic results emerge from the theory.  

\subsubsection{General theory in the presence of impurities}

It is assumed that layer $I\!I$ contains a certain amount of impurities. The Gibbs free energy of the system per unit area is given by
\begin{eqnarray}
  & & \hspace{-.5cm} G(T, P, N_s, N_l , N_i) \nonumber \\ &  = &  \mu_s(T,P) N_s + \mu_l(T,P) N_l + \mu_i(T,P) N_i \nonumber \\ & & + R_g T N_i \left[ \ln\frac{N_i}{N_l} - 1 \right] + \mathcal{I}(d) \, ,
\end{eqnarray}
where $T$ and $P$ are temperature and pressure, $\mu_s$, $\mu_l$, $\mu_i$ are the chemical potentials per mole of the solid, the liquid, and the impurity species, respectively, $N_s$, $N_l$, $N_i$ are the numbers of mole per unit area of the solid, the liquid, and the impurities, respectively. The term $R_g T N_i \left[ \ln\frac{N_i}{N_l} - 1 \right]$, where $R_g$ is the gas constant, corresponds to the mixing entropy of the impurities assuming low impurity concentration so that the ideal gas result applies. Finally, $\mathcal{I}(d)$ denotes the total effective interfacial free energy described in more detail below. Note that $d$ is directly proportional to $N_l$ through $N_l = \rho_l d$, where $\rho_l$ is the molar density of the liquid which is a constant in the calculation. 

The solid (ice) and the liquid (water) parts of the system can exchange material and hence the condition for thermodynamic equilibrium is given by
\begin{equation}
  \frac{\partial G}{\partial N_s} = \frac{\partial G}{\partial N_l} \, .
\end{equation}
Applying this criterion to the free energy given above, and using the fact that ${\partial \mu_{s,l}}/{\partial T} \equiv -s_{s,l}$ where $s_{s}$ ($s_{l}$) is the specific entropy of the solid (liquid), we can approximate $\mu_l - \mu_s \approx q_m \frac{\Delta T}{T_m}$, where $q_m$ is the latent heat of melting per mole and arrive at
\begin{equation}
\label{eq_geninterf}
   \rho_l q_m \frac{\Delta T}{T_m} = \frac{R_g T_m N_i}{d} - \mathcal{I}^{\prime}(d) \, .
\end{equation}
In the absence of interfacial forces we have $\mathcal{I}^{\prime}(d)=0$, and the result reduces to $\Delta T = \frac{R_gT_m^2}{q_m\rho_l} \rho_i$, where the molar density of impurities is $\rho_i = N_i/d$. This undercooling corresponds precisely to the classical colligative effect--the depression of the bulk freezing temperature due to dissolved impurities. Obviously,  compared to the force free case, repulsive interfacial forces for which $\mathcal{I}^{\prime}(d)<0$ result in a larger undercooling $\Delta T$ thereby enhancing the layer thickness. Conversely, attractive interfacial forces wherein $\mathcal{I}^{\prime}(d)>0$ suppress the formation of a premelted liquid film.  Indeed, for sufficiently large $\Delta T$, Eq.~\eqref{eq_geninterf} may have no solution showing that the attractive interfacial forces can dominate the colligative effect of the impurities and the premelted liquid layer will vanish.

\subsubsection{Contributions to the effective interfacial free energy $\mathcal{I}(d)$}

The interfacial energy has two contributions; $\mathcal{I}(d)= F_{\text{dis}}(d) + F_{\text{elec}}(d)$, where $F_{\text{dis}}(d)$ is due to dispersion forces, acting between the three interfaces defining the system, and $F_{\text{elec}}(d)$ captures the effect of immobile charges at the interface. Although we compute the strength of the dispersion force interaction from the full frequency dependent theory \cite{DzLiPi61}, we take the small thickness limit in which only non-retarded forces contribute to $F_{\text{dis}}(d)$ which consequently can be written as
\begin{equation}
 F_{\text{dis}}(d) = -\frac{A_H}{12\pi d^2} \, ,
\end{equation}
where $A_H$ is the Hamaker constant for the given layered system  \cite{WiWeElSch95, DRW06}. For symmetric systems, such as the ice-liquid-ice configuration, which occurs in the case of grain boundary melting, $A_H$ is {\em positive} and hence dispersion forces are {\em attractive}. At the interface between the ice and the porous matrix $A_H$ can assume both signs, depending on the material properties of the porous medium.

Immobile surface charges with a given surface charge density $q_s$ on both interfaces which bound the quasi-liquid layer are screened by the counterions in the liquid. Making use of the Debye-H{\"u}ckel theory this leads to the contribution
\begin{equation}
 F_{\text{elec}}(d) = \frac{2 q_s^2}{\kappa \epsilon \epsilon_0} e^{-\kappa d} \, ,
\end{equation}
where $\epsilon_0$ is the vacuum permittivity, $\epsilon$ the relative permittivity of liquid water, and
\begin{equation}
  \kappa^{-1} = \sqrt{\frac{\epsilon\epsilon_0 k_B T}{e^2 N_A \rho_i}} = \sqrt{\frac{\epsilon\epsilon_0 k_B T d}{e^2 N_A N_i}}
\end{equation}
is the Debye length for monovalent ions, with the electron charge $e$, Avogadro's number $N_A$, and Boltzmann's constant $k_B$. The Debye length describes the characteristic decay of the ion field adjacent to the charged surface due to the screening by counter ions. A repulsive force between two charged surfaces originates in the restriction of the entropy of the ions as the surfaces are brought closer.  Whereas ions of charge opposite to those of the surface are attracted to it, they are repelled from each other, and the
increased proximity induced by decreasing the film thickness increases the free energy.
Note the relatively complex $d$-dependence of $F_{\text{elec}}(d)$ which is due to the fact that $\kappa^{-1}$ is also a function of $d$.  Thus, the richness of the role of impurities in the premelting of ice is a consequence of this and the efficient segregation of ions into the film; since nearly all of the impurities remain in the liquid, as the temperature changes the {\em amplitude} and {\em range} of the repulsive electrostatic interaction varies \cite{We99}.  In principle, the situation could be even more complex when for example $q_s$ = $q_s(d)$.  Indeed, when $d$ is small the Debye-H{\"u}ckel expression is less accurate and steric effects become more important.  However, to introduce such complications here exceeds our understanding of the situation based on present experimental evidence.

Using these expressions for $\mathcal{I}(d)= F_{\text{dis}}(d) + F_{\text{elec}}(d)$ in Eq.~\eqref{eq_geninterf} we obtain an implicit equation for the thickness $d$ of the premelted layer as a function of the undercooling $\Delta T$:
\begin{equation}
\label{eq_explinterf}
   \rho_l q_m \frac{\Delta T}{T_m} = \frac{R_g T_m N_i}{d} - \frac{A_H}{6\pi d^3} + \frac{q_s^2}{\epsilon\epsilon_0} \left[ 1 - \frac{1}{c\sqrt{N_i d}}\right] e^{-c\sqrt{N_i d}}\, ,
\end{equation}
where the constant $c$ is
\begin{equation}
  c = \sqrt{\frac{e^2 N_A}{\epsilon\epsilon_0 k_B T_m}} \, .
\end{equation}
This result is analyzed in detail in Section~\ref{sec_resLayer}.

\subsection{Curvature induced melting: The Gibbs-Thomson Effect}

In order to assess the effect of the ice-liquid interfacial curvature on the freezing temperature, we neglect the known anisotropy of the surface energy of ice \cite{DRW06} and consider a spherical ice crystal of radius $r$ surrounded by undercooled water. As opposed to the planar case, the interfacial area is no longer a constant as it was in the treatment of Section~\ref{sec_mechLayer}.  With this new complication we reconsider the Gibbs free energy $G$ of the {\em total} system. The same holds for the numbers of moles $N_s$, $N_l$, and $N_i$ which are now taken for the total system and not per unit surface. For ease of  notation we use the same symbols for these new quantities. All other quantities are as defined above. Whence, we have
\begin{eqnarray}
   & & \hspace{-.5cm} G(T, P, N_s, N_l , N_i) \nonumber \\ & = & \mu_s(T,P) N_s + \mu_l(T,P) N_l + \mu_i(T,P) N_i \nonumber \\ & & + R_g T N_i \left[ \ln\frac{N_i}{N_l} - 1 \right] + 4\pi r^2 \gamma_{sl} \, ,
\end{eqnarray}
where $r$ is related to $N_s$ through $N_s = \frac{4\pi}{3} r^3 \rho_s$ with the constant molar density $\rho_s$. 
We note here that we understand that $\gamma_{sl} = \gamma_{sl} (r, N_i, \phi)$, where $\phi$ is the angle that defines the surface normal.  Thus, the surface energy, itself a Gibbs free energy conjugate to area, is known in general to depend on the size of the system, the adsorption of impurities and the crystallographic orientation \cite{DRW06} but the simplification of treating it as a constant is justified both on the basis of experimental evidence for the first two dependencies and on the level of our analysis.

The condition for thermodynamic equilibrium is again
\begin{equation}
  \frac{\partial G}{\partial N_s} = \frac{\partial G}{\partial N_l} \, ,
\end{equation}
and hence a calculation analogous to that in the Section~\ref{sec_mechLayer} yields the result
\begin{equation}
  q_m \frac{\Delta T}{T_m} - R_g T_m \frac{\rho_i}{\rho_l} = \frac{2\gamma_{sl}}{\rho_s r} \, ,
\end{equation} 
or, for the general case of a surface with principal radii of curvature $r_1$ and $r_2$,
\begin{equation}
\label{eq_gencurvmelt}
  q_m \frac{\Delta T}{T_m} - R_g T_m \frac{\rho_i}{\rho_l} = \frac{\gamma_{sl}}{\rho_s}\left(\frac{1}{r_1} + \frac{1}{r_2} \right) \, .
\end{equation} 
This result is the well known Gibbs-Thomson effect, i.e. the freezing point depression due to convexity of the ice-liquid interface, calculated here for the case where impurities are present in the liquid water. The larger the undercooling $\Delta T$, the smaller the radii of curvature of an ice-liquid interface in equilibrium. At the bulk freezing point, $\Delta T \searrow R_gT_m^2\rho_i/q_m\rho_l$, the radius of curvature diverges and a planar interface is the stable configuration.

\subsection{Premelting in a porous medium}

\label{sec_premeltPorous}

We use the random close packing (RCP) of hard spheres as a model for the porous matrix. Apart from choosing the matrix material and hence calculating the properties which enter the determination of the Hamaker constant (see Section~\ref{sec_mechLayer}), the only free parameter characterizing the porous matrix which contains the partially premelted ice is the radius $R$ of the hard spheres. The properties of RCP hard spheres which are required for our calculation are the sphere packing fraction $\eta$ and the average coordination number $\mathcal{C}$. Recent estimates of these numbers \cite{SoWaMa08} are  $\eta = 0.634$ and $\mathcal{C} = 6$. It should be noted, however, that the basic concept of RCP is not well defined mathematically, so these numbers vary slightly according to the protocol and algorithm  used to determine the exact configuration \cite{ToTrDe00}.

At a given undercooling $\Delta T$ we are interested in the fraction of the interstitial volume which is occupied by quasi-liquid water $f_l$ which we will more compactly refer to as the liquid fraction.  The first contribution we consider comes from interfacial premelting at the surface of the porous matrix. In the limit where the premelted layer is much thinner than the radius of the matrix particles, $d\ll R$, this fraction is well approximated by
\begin{equation}
  f_l^{\text{mat}}(\Delta T, \rho_i) = \frac{ a_{\text{mat}}}{1-\eta}\times d(\Delta T, \rho_i ; A_H^{\text{mat}}, q_s^{\text{mat}}) \, ,
\end{equation}
where $a_{\text{mat}}$ is the surface area of the matrix particles per unit volume and $d$ is obtained from Eq.~\eqref{eq_explinterf} using the Hamaker constant $A_H^{\text{mat}}$ for the ice-water-matrix layered system and the appropriate surface charge density $q_s^{\text{mat}}$. The sphere number density of the matrix is ${n_s}=\eta/(4\pi R^3/3)$ and thus we obtain $a_{\text{mat}} = 4\pi R^2 {n_s} = 3\eta/R$ and 
\begin{equation}
\label{eq_flmatlayer}
  f_l^{\text{mat}}(\Delta T, \rho_i) = \frac{ 3\eta }{1-\eta} \times d(\Delta T, \rho_i ; A_H^{\text{mat}}, q_s^{\text{mat}}) R^{-1} \, .
\end{equation}

The second contribution to the liquid fraction is due to grain boundary melting. Its importance depends sensitively on the grain size of the interstitial ice, which is a function of such factors as (1) the rate of cooling, (2) the undercooling $\Delta T$, and (3) the time allowed for annealing.  Even were we to attempt a complete computation of the coarsening rate of the ice grains in the matrix, there is no unique way to specify the initial value problem and hence to arrive at a concrete grain size as a function of the properties of the porous matrix.  Thus, we introduce as an additional parameter the grain boundary density $\rho_{gb}$,  which is the surface area of grain boundaries per unit volume of the interstitial ice. We denote $\tilde{\rho}_{gb}=\rho_{gb}R$ as the related dimensionless density. The corresponding liquid fraction follows immediately
\begin{equation}
\label{eq_gblayer}
  f_l^{\text{gblayer}}(\Delta T, \rho_i) = \tilde{\rho}_{gb} \times d(\Delta T, \rho_i ; A_H^{\text{gb}}, q_s^{\text{gb}}) R^{-1}\, ,
\end{equation}
where $A_H^{\text{gb}}$ is the Hamaker constant for the ice-liquid-ice layered system and $q_s^{\text{gb}}$ is the surface charge density for the grain boundary. The thickness $d$ of the quasi-liquid layer is again obtained from Eq.~\eqref{eq_explinterf}.

\begin{figure}[t]

\begin{center}

\includegraphics[width=.45\textwidth]{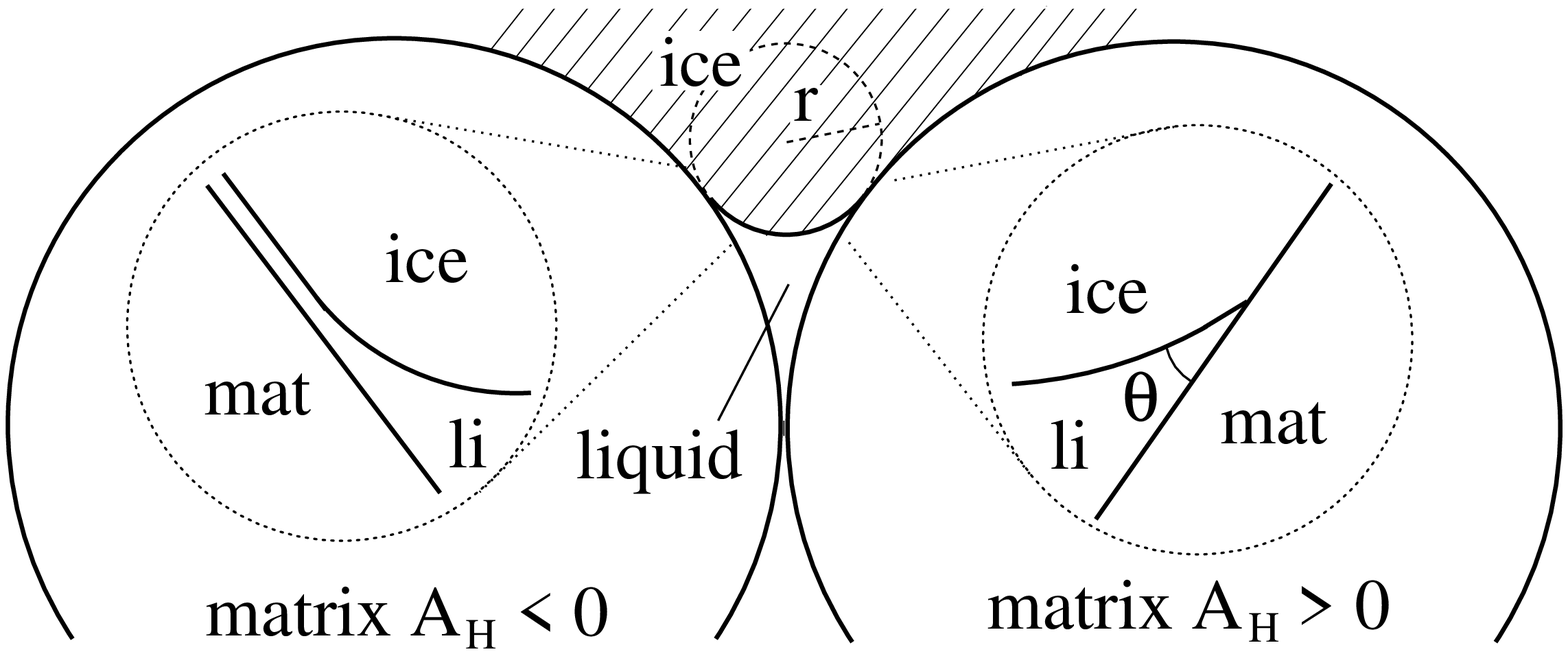}

\caption{Sketch of the liquid pocket which forms between two matrix spheres of radius $R$. The amount of premelted liquid depends on the characteristic radius of curvature $r$ of the ice-liquid interface as obtained from Eq.~\eqref{eq_rcurvmelt}. Depending on the matrix material there is either always a premelted liquid layer (repulsive dispersion forces, $A_H<0$) or the premelted layer may collapse under certain conditions (attractive dispersion forces, $A_H>0$) and the liquid-ice interface forms an angle $\theta < \pi/2$ with the matrix surface. For parsimony of presentation we show the neighboring matrix spheres of different materials while the present theory assumes porous media made up of a single material.}

\label{fig_contactPoint}

\end{center}

\end{figure}

We now deal with curvature induced premelting. Important contributions to the liquid water fraction come from regions where the ice-liquid interface has a large curvature. In our idealized matrix these are the points where two matrix spheres are in contact. Following CDF \cite{CaDaFu92} there is a liquid pocket forming at the contact point with a volume $2\pi r^2 R$ to lowest order in $r$, which is the radius of curvature of the ice-liquid interface (Fig.~\ref{fig_contactPoint}). Taking into account that the second principal radius of curvature is much larger than $r$, we can use Eq.~\eqref{eq_gencurvmelt} to relate $r$ to the undercooling $\Delta T$:
\begin{equation}
\label{eq_rcurvmelt}
  r(\Delta T, \rho_i) = \frac{\gamma_{sl}}{\rho_s} \left(q_m \frac{\Delta T}{T_m} - R_g T_m \frac{\rho_i}{\rho_l}\right)^{-1} \, .
\end{equation}
Using the number density ${n_s}$ of the sphere matrix and the matrix coordination number $\mathcal{C}$ we can express the corresponding fraction of liquid water with respect to the total interstitial volume:
\begin{eqnarray}
  f_l^{\text{contact}}(\Delta T, \rho_i) & = & \frac{\mathcal{C}/2 \times {n_s} \times 2\pi [r(\Delta T, \rho_i)]^2 R}{1-\eta} \nonumber \\
                         & = & \frac{3}{2} \times \frac{\eta}{1-\eta} \times \frac{\mathcal{C}}{2} \times \left[\frac{r(\Delta T, \rho_i)}{R}\right]^2 \, , \nonumber \\
\end{eqnarray}
with $r(\Delta T, \rho_i)$ from Eq.~\eqref{eq_rcurvmelt}.

A further curvature induced contribution to the liquid fraction brought out by CDF  comes from the lines where a grain boundary and a matrix sphere meet.

In the limit that the matrix particles are small on the scale of the linear scale of the ice grains, but that $R$ is much larger than $r(\Delta T, \rho_i)$ the corresponding liquid fraction can be calculated using elementary geometry (see Appendix~\ref{app_grainbmat})
\begin{eqnarray}
 && f_l^{\text{gbmat}}(\Delta T, \rho_i) \nonumber \\ && = 3\left(2-\frac{\pi^2}{8}\right) \times \frac{\eta}{1-\eta} \times \tilde{\rho}_{gb} \times \left[\frac{r(\Delta T, \rho_i)}{R}\right]^2 . \quad
\end{eqnarray}

Finally, curvature induced premelting gives rise to liquid veins along the lines where several grain boundaries meet. With the simplifying assumption that the ice grains are arranged in a cubic packing, the total length of veins per interstitial volume, or the density of veins $\rho_{\text{veins}}$,  is obtained from $\rho_{\text{gb}}$ as $\rho_{\text{veins}} = \rho_{\text{gb}}^2/3$. Given that the cross-sectional area of a vein in cubic packing is $(4-\pi)r^2$, we can calculate the corresponding contribution to the liquid fraction as
\begin{eqnarray}
  f_l^{\text{veins}}(\Delta T, \rho_i) & = & \rho_{\text{veins}}\times (4-\pi) \times [r(\Delta T, \rho_i)]^2 \nonumber \\
                         & = & \frac{1}{3}(4-\pi) \times \tilde{\rho}_{gb}^2 \times  \left(\frac{r(\Delta T, \rho_i)}{R}\right)^2  . \qquad 
\end{eqnarray}

Therefore, the total liquid fraction  is obtained as the sum of the five individual contributions which we have derived above; $f_l^{\text{mat}}, f_l^{\text{gblayer}}, f_l^{\text{contact}}, f_l^{\text{gbmat}}$, and $f_l^{\text{veins}}$.

\subsection{Implicit equation for the total liquid fraction}

\label{sec_implLiquidFr}

Thus far we have taken impurities into account in terms of a molar density $\rho_i$ or, in the case of the premelted layer, a surface density of impurities $N_i = \rho_i d$. Laboratory probes using scattering or related approaches involve relatively small samples with good thermal control and a well defined (fixed for a given run) dopant level \cite{DRW06}.  The temperature changes in these systems are quasi-static so that the sample is at coexistence.  In the natural environment there are of course temperature gradients.  Therefore, for applicability of our work there, the typical broad temperature gradients found for example in natural soils imply that a local region within will be nearly isothermal and thus isosaline.  Clearly, there will be many environmental circumstances where this may be violated, but because the first test of such a theory should be of high precision we focus mainly on the simplest experimental setting and exploit the fact that the impurity segregation coefficient of ice is small \cite{GrGuCa87} so that the premelted liquid is enriched/diluted when the system is cooled/warmed. Hence it is prudent to introduce $\rho_0$, the molar density of impurities in the interstitial water when the system is above the bulk melting temperature. 
Thus, because of the validity of the assumption that impurities are contained almost exclusively in the liquid fraction of the interstitial water it follows that
\begin{equation}
  \rho_i = \frac{\rho_0}{f_l} \, ,
\end{equation}
thereby properly accounting for the increase in the impurity concentration in the premelted liquid as the liquid fraction decreases. Due to the fact that the regions of premelted water are all interconnected, $\rho_i$ is homogeneous over the entire sample.

Having now considered the essential physical phenomena we can formulate an equation for the total liquid fraction $f_l^{\text{tot}}$ in the porous medium as a function of the undercooling $\Delta T$ and the impurity concentration $\rho_0$ in the completely melted sample. Combining all the contributions discussed in Section~\ref{sec_premeltPorous}, we obtain
\begin{eqnarray}
\label{eq_fltot}
 f_l^{\text{tot}} &  = & f_l^{\text{mat}}(\Delta T, \rho_0/f_l^{\text{tot}}) + f_l^{\text{gblayer}}(\Delta T,  \rho_0/f_l^{\text{tot}})  \nonumber \\
  &  & + f_l^{\text{contact}}(\Delta T,  \rho_0/f_l^{\text{tot}}) + f_l^{\text{gbmat}}(\Delta T,  \rho_0/f_l^{\text{tot}}) \nonumber \\ & & + f_l^{\text{veins}}(\Delta T,  \rho_0/f_l^{\text{tot}}) \label{eq_implfl} \, 
\end{eqnarray}
which is an implicit function for $f_l^{\text{tot}}=f_l^{\text{tot}}(\Delta T, \rho_0)$ and in the general case has no analytical solution. Moreover, the underlying equation \eqref{eq_explinterf} for the thickness $d$ of the premelted layer, which is required for $f_l^{\text{mat}}$ and $f_l^{\text{gblayer}}$, is itself an implicit function of $d$. This calls for a careful stepwise analysis first of Eq.~\eqref{eq_explinterf} and subsequently of Eq.~\eqref{eq_implfl}, which we perform in Sections~\ref{sec_resLayer} and \ref{sec_resLiquidFr}.

\section{Analysis and application}

\label{sec_analysis}

\subsection{Thickness of the premelted layer}
\label{sec_resLayer}

In Section~\ref{sec_mechLayer} we derived Eq.~\eqref{eq_explinterf} for the thickness $d$ of the premelted  layer  as a function of the undercooling $\Delta T$, the impurity concentration $N_i$, the dispersion forces as embodied in the Hamaker constant $A_H$, and the surface density $q_s$ of immobile charges on the solid-liquid interfaces.  This expression cannot be solved analytically in the general case.  Hence, presently we analyze 
Eq.~\eqref{eq_explinterf} analytically, to demonstrate that we can obtain known distinguished limits and facilitate intuition, and numerically, to provide specific ranges of results for particular physical systems of interest.  We begin by rewriting this equation in terms of the reduced, or dimensionless, undercooling $t=\Delta T/T_m$ as follows
\begin{equation}
\label{eq_filmnodim}
 t = \frac{b N_i}{d} - \frac{a}{d^3} + q \left( 1 - \frac{1}{c\sqrt{N_i d}}\right) e^{-c\sqrt{N_i d}} \, ,
\end{equation}
where the constants $b = R_g T_m/(\rho_l q_m)$, $a = A_H/(6\pi\rho_l q_m)$, and $q=q_s^2/(\epsilon\epsilon_0\rho_l q_m)$. For {\em attractive} dispersion forces which {\em suppress} premelting we have $a>0$, and for {\em repulsive} dispersion forces which {\em promote} premelting we have $a<0$.  All other quantities in Eq.~\eqref{eq_filmnodim} are positive. 

\subsubsection{Maximal undercooling and minimal layer thickness}

The simplest case occurs when both dispersion and electrostatic forces are absent, i.e., $a=0$ and $q=0$. The thickness $d$ of the liquid film is them simply given by
\begin{equation}
\label{eq_colligativeOnly}
  d = {b N_i}{t}^{-1} \, ,
\end{equation}
representing the colligative effect of the dissolved impurities which, due to the small segregation coefficient of ice, are strongly enriched in the liquid layer. The situation immediately becomes more complex if we take dispersion forces into account. Let us first consider materials that have $a>0$, which means that dispersion forces suppress the formation of a premelted liquid film. Assuming $q=0$ for the moment, Eq.~\eqref{eq_filmnodim} possesses a solution as long as the undercooling $t$ does not exceed a critical value $t_{\text{max}} =   \frac{2}{3}  \frac{(b N_i)^{3/2}}{\sqrt{3a}}$ corresponding to a minimum film thickness $d_{\text{min}} = \sqrt{\frac{3a}{bN_i}}$. If the temperature is reduced such that $t>t_{\text{min}}$ the attractive dispersion forces dominate the colligative effect and the liquid film collapses to zero thickness.

\begin{figure}[t]

\begin{center}

\includegraphics[width=.45\textwidth]{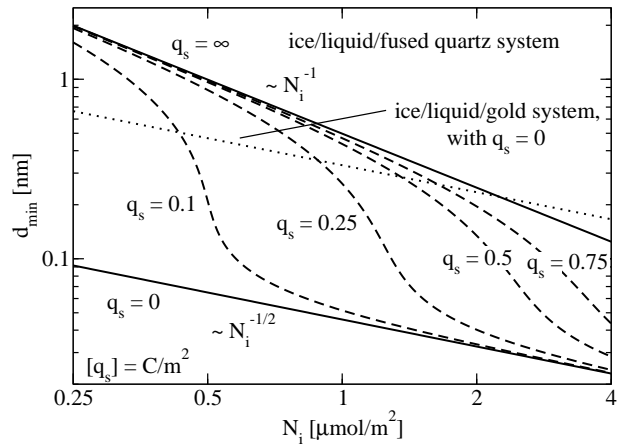}

\caption{In the case where dispersion forces at the interface are attractive, the quasi-liquid layer always collapses discontinuously from a minimal thickness $d_{\text{min}}$, plotted here as a function of the impurity density $N_i$ for different values of the surface charge $q_s$,  to zero when the temperature is lowered such that the undercooling exceeds $\Delta T_{\text{max}}$ (plotted in Fig.~\ref{fig_tmax}). The collapse is caused by the combined effect of immobile surface charges (dominating as $N_i\to 0$) and dispersion forces (dominating as $N_i\to\infty$). For the ice/liquid/fused quartz system the collapse caused be dispersion forces occurs at thicknesses on the scale of the size of a single molecule where the notion of a quasi-liquid layer no longer applies. In order to observe collapse induced by dispersion forces a larger Hamaker constant is required such as provided by the ice/liquid/gold system.}

\label{fig_dmin}

\end{center}

\end{figure}

\begin{figure}[t]

\begin{center}

\includegraphics[width=.45\textwidth]{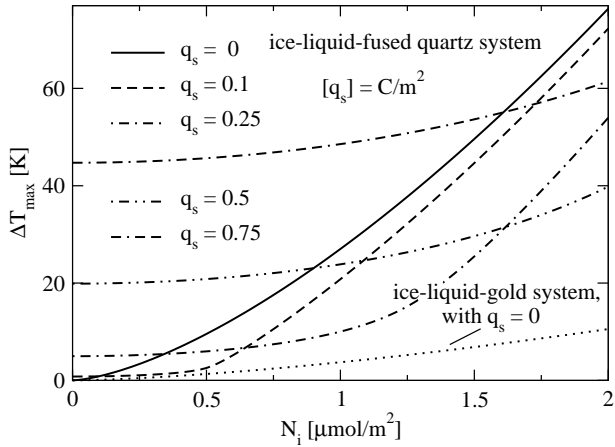}

\caption{The maximal undercooling $\Delta T_{\text{max}}$ as a function of the impurity concentration $N_i$ for the same parameters as used for the calculation of $d_{\text{min}}$ in Fig.~\ref{fig_dmin}. The zero slope asymptote for $N_i\to 0$ is observed as expected from Eq.~\eqref{eq_dminTmaxNiZero}. The large Hamaker constant of the ice-liquid-gold system leads to low values of $\Delta T_{\text{max}}$.}

\label{fig_tmax}

\end{center}

\end{figure}

For the most general form of Eq.~\eqref{eq_filmnodim}, the maximal undercooling $t_{\text{max}}$ and corresponding minimal layer thickness $d_{\text{min}}$ at which the quasi-liquid layer collapses cannot be expressed analytically. We obtain the global maximum $(d_{\text{min}},t_{\text{max}})$ by extremizing the right hand side of Eq.~\eqref{eq_filmnodim} with respect to $d$.  Two limiting cases can be distinguished. For large impurity concentrations $N_i$ the contribution due to surface charges becomes negligible and hence we recover the result
\begin{eqnarray}
   d_{\text{min}}(N_i\to\infty) & \to & \sqrt{\frac{3a}{bN_i}} \, , \nonumber \\  t_{\text{max}}(N_i\to\infty) & \to & \frac{2}{3}  \frac{(b N_i)^{3/2}}{\sqrt{3a}} \, , \label{eq_dminTmaxNiInfty}
\end{eqnarray}
 for the case $q=0$ which we derived above. On the other hand, it can be shown that in the limiting case $N_i\to 0$ the values $d_{\text{min}}$ and $t_{\text{max}}$ follow from the surface charge dominated limit ($b=0$ and $a=0$) which has
\begin{equation}
\label{eq_dminTmaxNiZero}
   d_{\text{min}}(N_i\to 0) \to \frac{2.62}{c^2 N_i} \, , \qquad t_{\text{max}}(N_i\to 0) \to 0.0757 \times q \, .
\end{equation}
In Fig.~\ref{fig_dmin} we plot the minimal thickness $d_{\text{min}}$ of the quasi-liquid layer corresponding to the maximal undercooling $t_{\text{max}}$ as a function of the impurity concentration $N_i$ for different values of the surface charge $q_s$. We use a Hamaker constant $A_H = 3.0\times 10^{-23}\text{ J}$ which is the result obtained by Wilen et al.\ \cite{WiWeElSch95} for the ice/liquid/fused quartz layered system. The analytical results for $N_i\to 0$ and $N_i\to\infty$ are 
characterized by the $N_i$-dependencies of  the asymptotic behavior.  Note that the minimal thickness in the dispersion force dominated limit ($q_s = 0$) is below $0.1\text{ nm}$ and therefore equal to the dimension of a single molecule. Hence, the notion of a quasi-liquid layer can no longer be applied and for the ice/liquid/fused quartz system layer collapse can only be observed experimentally at   low impurity concentrations and it is always triggered by surface charges. This is seen directly from Eq.~\eqref{eq_filmnodim} where the contribution of the surface charge  for sufficiently small $d$ becomes {\em negative}, indicating {\em attractive} forces, which suppress liquid layer formation. In order to observe a collapse of the quasi-liquid layer driven by dispersion forces, the latter have to be more strongly attractive. We have included in Fig.~\ref{fig_dmin} the case of the ice-liquid-gold system which has a Hamaker constant of $A_H = 1.573\times 10^{-21}\text{ J}$ \cite{WiWeElSch95}, and hence dispersion forces are indeed strong enough to yield a liquid-layer collapse at thicknesses of the order of those calculated for the surface charge dominated case. The contrast between the two scenarios will be studied further in the second part of this section (Figs.~\ref{fig_layerthicknessQuartz} and \ref{fig_layerthicknessGold}). In Fig.~\ref{fig_tmax} the corresponding results for the maximal undercooling $\Delta T_{\text{max}} = T_mt_{\text{max}}$ are plotted. As we expect from Eqs.~\eqref{eq_dminTmaxNiInfty} and \eqref{eq_dminTmaxNiZero}, $\Delta T_{\text{max}}$ increases with surface charge $q_s$  resulting in a $q_s$-dependent offset for $N_i\to 0$. Moreover, for all $q_s$, $\Delta T_{\text{max}}$ increases with impurity concentration $N_i$.  A large Hamaker constant, for example of the ice/liquid/gold system, results in a low $\Delta T_{\text{max}}$ as shown by Eq.~\eqref{eq_dminTmaxNiInfty}, where $\Delta T_{\text{max}} \propto A_H^{-1/2}$.  There are interesting experimental consequences of Fig.~\ref{fig_tmax}. Obviously, measuring $\Delta T_{\text{max}}$ for just one given impurity concentration $N_i$ is {\em not} sufficient to determine the surface charge $q_s$ at the interface. This follows from the fact that any two curves intersect, and hence measurements for at least two different impurity concentrations are required.

\subsubsection{Layer thickness for a given undercooling}

When $q=0$, Eq.~\eqref{eq_filmnodim} can still be solved analytically but because the resulting expression is lengthy and not particularly instructive we present the solution only in the small $t$ limit of moderate undercooling, 
\begin{equation}
\label{eq_limitLowDTnoQ}
 d = \frac{b N_i}{t} - \frac{a t}{b^2 N_i^2} - \mathcal{O}(t^3) \, .
\end{equation}
This result clearly illustrates the fact that the presence of attractive dispersion forces ($a>0$) acting across the system suppresses the formation of a quasi-liquid layer, whereas repulsive dispersion forces ($a<0$) tend to increase the thickness of the premelted layer.

When surfaces charges on the interfaces of the quasi-liquid layer with the substrate/ice are involved, Eq.~\eqref{eq_filmnodim} becomes transcendental and has no analytical solution. In order to get some insight in the effect of surface charges, we consider several special cases. First, let us assume that for moderate undercooling the system is {\em dominated} by surface charges. This scenario corresponds to setting $a=0$, $b=0$ in Eq.~\eqref{eq_filmnodim} and considering only the leading contributions as $d$ approaches infinity. The solution is then
\begin{equation}
\label{eq_chargeDom}
  d = \frac{[\ln (\frac{q}{t})]^2}{c^2 N_i} \, .
\end{equation}
This result reveals an interesting consequence of the presence of charges at the interface. Under appropriate conditions the layer thickness $d$ {\em decreases} when the impurity concentration {\em increases}. This observation is contrary to the naive expectation based on the colligative effect and originates in the fact that for larger $N_i$ the electrostatic repulsion is screened for effectively \cite{We99}.

For repulsive dispersion forces ($a<0$) the solution of Eq.~\eqref{eq_filmnodim} can be obtained asymptotically in the limit of large undercooling $t\to\infty$, whereas for attractive dispersion forces ($a>0$)  the premelted layer collapses to zero thickness for $t$ larger than some finite value. The repulsive case is
\begin{equation}
\label{eq_largeDT}
 d(t, N_i; a, q) = \frac{(-a)^{1/3}}{t^{1/3}} + \frac{b N_i}{3t} - \frac{(-a)^{1/6} q}{3 c \sqrt{N_i} t^{7/6}} + \mathcal{O}\left(\frac{1}{t^{4/3}}\right) \, 
\end{equation}
and seems to contradict Eq.~\eqref{eq_chargeDom} for two reasons. First,  the negative sign of the term proportional to $q$ in Eq.~\eqref{eq_largeDT} indicates that immobile surface charges impede the formation of the premelted layer, whereas in  Eq.~\eqref{eq_chargeDom} immobile surface charges alone are sufficient to stabilize a premelted layer. Second, while Eq.~\eqref{eq_chargeDom} implies that $d$ decreases when the impurity concentration $N_i$ is increased, the opposite conclusion follows from Eq.~\eqref{eq_largeDT} where, due to the term $\propto q$, the inhibition due to surfaces charges is reduced when $N_i$ increases.

The apparent contradiction can be reconciled by the following observations. A direct consequence of Eq.~\eqref{eq_filmnodim} is that immobile surface charges do not contribute to the film thickness $d$ when the latter equals the Debye length, namely $d_0=\frac{1}{c^2 N_i}$, in which case the surface charge term vanishes. The film thickness $d_0$ translates into an undercooling $t_0 = b c^2 N_i^2 - a c^6 N_i^3$. For an undercooling $t<t_0$ surface charges enhance the formation of a quasi-liquid layer while for $t>t_0$ surface charges suppress layer formation.  Considering that Eq.~\eqref{eq_chargeDom} applies to the $t\to 0$ case while Eq.~\eqref{eq_largeDT} to the $t\to\infty$ case, this explains the essential difference between Eqs.~\eqref{eq_chargeDom} and \eqref{eq_largeDT} in terms of whether they support or suppress the formation of a liquid film.

In order to isolate the dependence of $d$ on the impurity concentration $N_i$ we consider Eq.~\eqref{eq_filmnodim} in the case where dispersion forces can be neglected ($a=0$) and the immobile interfacial charge density is small (i.e.\ $q \to 0$). To lowest order $d$ depends linearly on $q$, and hence we obtain
\begin{eqnarray}
  && d(t, N_i; a=0, q\to 0) \nonumber \\ && = \frac{b N_i}{t} + \frac{b q}{t^2}\left(N_i - \frac{\sqrt{t}}{c\sqrt{b} }\right) \times e^{-c N_i \sqrt{b/t}} + \mathcal{O}(q^2) \, .\qquad \label{eq_lowQExp}
\end{eqnarray}
From this result the sign of $\partial d / \partial N_i$ can be shown to equal the sign of
\begin{equation}
\label{eq_alphabeta}
 {\cal S}\equiv \alpha^2(2-\alpha) + e^{\alpha}/\beta \, ,
\end{equation}
where $\alpha = c N_i \sqrt{b/t}$ and $\beta = q/(bc^2N_i^2)$. We find ${\cal S}>0$ for all $\alpha>0$ as long as $\beta < \beta_c = e^4/32 \simeq 1.71$. Therefore if $N_i>\sqrt{q/(\beta_c b c^2)}$ the layer thickness $d$ {\em increases} with $N_i$ for any given undercooling $t$. For lower impurity concentrations ($\beta>\beta_c$) there is a finite interval for $\alpha$ which contains $\alpha_c = 4$, where ${\cal S}<0$. Therefore, at sufficiently low impurity concentration there exists an interval of intermediate undercooling $t$ where $d$ {\em decreases} when $N_i$ increases. This result implies that for large undercooling $t$ the layer thickness $d$ must always increase with $N_i$ which was precisely a conclusion obtained from Eq.~\eqref{eq_largeDT}. However, the result cannot be applied to the charge dominated case, Eq.~\eqref{eq_chargeDom}, because the corresponding limit $b\to0$ renders Eq.~\eqref{eq_lowQExp} trivial. It is  possible though to obtain ${\cal S}<0$ in the related limit $b\to 0$, $t\to0$, if we assume that $b/t$ approaches a large constant such that $\alpha >2$ holds.

\paragraph*{Attractive dispersion forces}

\begin{figure}[t]

\begin{center}

\includegraphics[width=.45\textwidth]{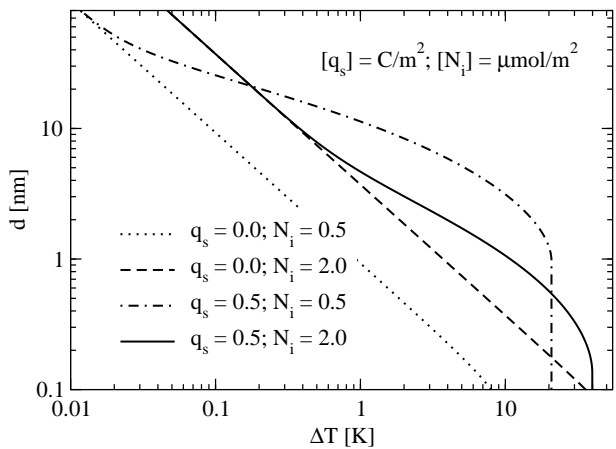}

\caption{Thickness $d$ of the quasi-liquid layer for the ice/liquid/fused quartz system ($A_H = 3.0\times 10^{-23}\text{ J}$) as a function of the undercooling $\Delta T$. For moderate undercooling the system is dominated by the colligative effect. In this regime the layer thickness increases with impurity density $N_i$. At intermediate undercooling the effect of immobile surface charges leads to a thickening of the quasi-liquid layer which is most effective for low impurity concentrations leading to a {\em decrease} of the layer thickness when the impurity density is increased. At large undercooling surface charge induced layer collapse is observed, while layer collapse due to dispersion forces alone occurs at layer thicknesses below $0.1\text{ nm}$.}

\label{fig_layerthicknessQuartz}

\end{center}

\end{figure}

\begin{figure}[t]

\begin{center}

\includegraphics[width=.45\textwidth]{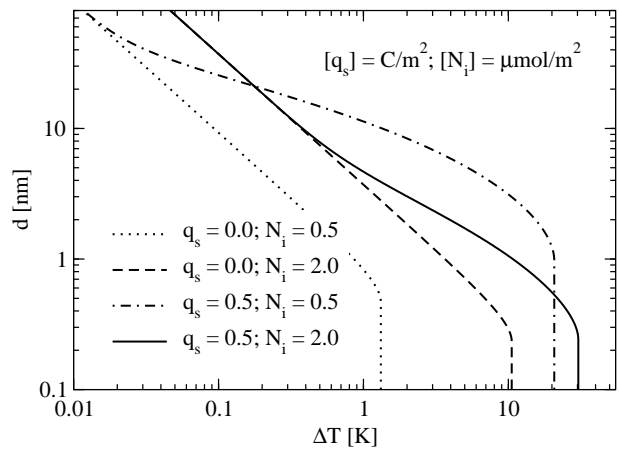}

\caption{Thickness of the quasi-liquid layer for the ice/liquid/gold system ($A_H = 1.573\times 10^{-21}\text{ J}$). In contrast to the ice/liquid/fused quartz system, Fig.~\ref{fig_layerthicknessQuartz}, layer collapse occurs at thicknesses above $0.1\text{ nm}$ even for the cases without surface charges. This is due to the stronger attractive dispersion forces in the ice/liquid/gold system. Layer collapse for the cases with surface charges is altered only for the higher impurity density $N_i = 2.0\text{ $\mu$mol/m$^2$}$ relative to the ice/liquid/fused quartz system, because at $d\approx 1\text{ nm}$ dispersion forces are weak compared to the electrostatic force.}

\label{fig_layerthicknessGold}

\end{center}

\end{figure}

The non-monotonic dependence of $d$ on $N_i$ is illustrated in Fig.~\ref{fig_layerthicknessQuartz} where the full numerical solution of Eq.~\eqref{eq_filmnodim} is plotted for the case of the ice-liquid-fused quartz layered system. At low undercooling $\Delta T$ the system displays characteristic  colligative behavior as described by Eq.~\eqref{eq_colligativeOnly}. The effect of immobile surface charges vanishes in the limit $t\to 0$ as indicated by Eq.~\eqref{eq_lowQExp} and the layer thickness $d$ increases when the impurity density $N_i$ increases. The layer thickness for small $\Delta T$ is obtained by requiring that the impurity {\em concentration} $\rho_i = N_i/d$ equals the impurity concentration which would be necessary in a bulk system to see a freezing point depression equal to the undercooling. When $\Delta T$ is increased, surfaces charges start to {\em increase} the layer thickness relative to the system without surface charges ($q_s = 0$). Interestingly, the resulting layer thickness for the system with the lower impurity density ($N_i = 0.5\text{ $\mu$mol/m$^2$}$) becomes even larger than for the system with more impurities ($N_i = 2.0\text{ $\mu$mol/m$^2$}$). This illustrates the result from the above analysis displaying the existence of an interval of intermediate undercooling where the layer thickness decreases when the impurity density increases. Recall that the reason for this is that at low impurity concentrations the electrostatic force is less effectively screened and hence the repulsion due to surface charges has a stronger effect, leading to a larger layer thickness than for higher impurity concentrations. When the system is undercooled further the collapse of the quasi-liquid layer can be observed in the figure. Here, because of the weakness of dispersion forces, the collapse is a consequence of the electrostatic force becoming attractive at low layer thicknesses. Collapse for the cases without surface charges cannot be observed in Fig.~\ref{fig_layerthicknessQuartz} as it occurs at thicknesses below $0.1\text{ nm}$ (c.f., Fig.~\ref{fig_dmin}). In order to see  layer collapse at larger layer thicknesses a larger Hamaker constant is required, as provided by the ice-liquid-gold system. To see this, we plot solutions of Eq.~\eqref{eq_filmnodim} for this system in Fig.~\ref{fig_layerthicknessGold} for the same parameters as considered in Fig.~\ref{fig_layerthicknessQuartz} for the ice-liquid-fused quartz system. Layer collapse is now observed at thicknesses above $0.1\text{ nm}$ even for the configurations without surface charges. For the configurations with surface charges we observe that for $N_i = 0.5\text{ $\mu$mol/m$^2$}$ layer collapse is not altered by the stronger dispersion forces while for $N_i = 2.0\text{ $\mu$mol/m$^2$}$ the stronger attractive dispersion forces lead to a layer collapse at a larger layer thickness compared to the corresponding configuration considered in Fig.~\ref{fig_layerthicknessQuartz}. Thus, in this case stronger dispersion forces drive layer collapse. 

\paragraph*{Repulsive dispersion forces}

\begin{figure}[t]

\begin{center}

\includegraphics[width=.45\textwidth]{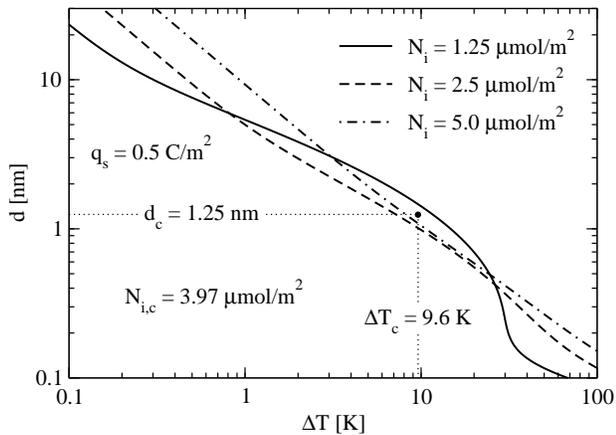}

\caption{Thickness $d$ of the quasi-liquid layer for the ice/liquid/silicon system ($A_H = -~1.66\times 10^{-21}\text{ J}$) as a function of the undercooling $\Delta T$. As dispersion forces in this system are repulsive they stabilize thin quasi-liquid layers and hence no layer collapse is observed. The figure illustrates the analytical result obtained from Eq.~\eqref{eq_lowQExp} which states that for intermediate undercooling $\Delta T$ and $N_i$ below a transition value the layer thickness {\em decreases} when the impurity concentration {\em increases}. For large $N_i$ the colligative effect dominates and $d$ increases when $N_i$ is increased. The transition value $N_{i,c}$ as well as the locus $\Delta T_c$ and $d_c$ where the interval with $\partial d/ \partial N_i < 0$ appears as derived from the figure are in good agreement with the analytical result (values given in the figure).  Note the sigmoidal transition to small but finite film thickness for the lowest value of $N_i$.}

\label{fig_layerthicknessSilicon}

\end{center}

\end{figure}

The result of the analysis following Eq.~\eqref{eq_lowQExp} can be tested on the ice/liquid/silicon system which has a Hamaker constant of $A_H = -1.66\times 10^{-21}\text{ J}$ \cite{WiWeElSch95}, the negative sign implying repulsive dispersion forces. As shown in Fig.~\ref{fig_layerthicknessSilicon} the system displays the expected colligative behavior for low undercooling and an enhanced layer thickness for intermediate undercooling where the effect of surface charges enhances the colligative effect, which alone would yield a straight line, $d\propto N_i^{-1}$. For intermediate undercooling the figure reveals a transition from a regime of low impurity density where the layer thickness {\em decreases} when $N_i$  {\em increases}, to a regime of high impurity density where the colligative effect dominates and accordingly  an {\em increase} of $d$ is observed when $N_i$  {\em increases}. From the figure the locus of the transition can be estimated to be $\Delta T_c\approx 10-20\text{ K}$, $d_c \approx 1\text{ nm}$, and $N_{i,c} \approx 2.5\text{ $\mu$mol/m$^2$}$. This is in good agreement with the analytical result obtained from Eq.~\eqref{eq_lowQExp}, $\Delta T_c = 9.6\text{ K}$, $d_c = 1.25\text{ nm}$, and $N_{i,c} = 3.97\text{ $\mu$mol/m$^2$}$. It should be noted that the analytical result is exact only in the limit $q_s\to 0$ but the complete solution of Eq.~\eqref{eq_filmnodim} shows that it applies qualitatively and semi-quantitatively for surface charge densities as high as $q_s = 0.5\text{ C/m$^2$}$. For the ice/liquid/silicon system no layer collapse is observed. This results from the fact that dispersion forces, which dominate at low layer thicknesses, are repulsive and stabilize the quasi-liquid layer. However, at impurity densities $N_i$ smaller than those considered in Fig.~\ref{fig_layerthicknessSilicon}, the Debye-H\"uckel expression for $F_{\text{elec}}$ generates an {\em attractive} force in the ice/liquid/silicon system with $q_s=0.5\text{ C/m$^2$}$ at sufficiently small layer thicknesses which results in a discontinuous drop in the layer thickness at a certain undercooling. The layer, however, does not collapse to zero thickness but rather is stabilized at a smaller thickness due to the repulsive dispersion forces which dominate for very thin quasi-liquid layers. In this article we do not investigate this phenomenon of layer discontinuity which occurs in a regime where both the $d$-dependence of $q_s$, the peculiarities of the Debye-H\"uckel result, and other effects,  have to be accounted for.

\subsection{Volume fraction of premelted water}
\label{sec_resLiquidFr}

Having analyzed premelting in planar  layered systems, we now return to the implicit equation (Eq.~\ref{eq_implfl}) for the volume fraction of premelted liquid at a given undercooling $t$ that we introduced in Section~\ref{sec_implLiquidFr}. In addition to being implicit for $f_l$ the equation uses the thickness of the premelted layer, which itself is obtained from an equation which does not have an analytical solution (Eq.~\ref{eq_filmnodim}). Accordingly, we do not possess an analytical expression for $f_l$ in the general case. 

\paragraph*{Low surface charge}

However, where immobile surface charges can be neglected ($q_s=0$), Eq.~\eqref{eq_implfl} becomes more tractable. In the case where the matrix material has a negative Hamaker constant ($A_H^{\text{mat}}<0$), i.e., dispersion forces are repulsive and support the formation of a premelted layer, Eq.~\eqref{eq_implfl} becomes
\begin{equation}
\label{eq_implflNoQAneg}
  f_l = \frac{A}{(t-c_0/f_l)^{1/3}} + \frac{B}{(t-c_0/f_l)^2} \, , 
\end{equation}
where
\begin{eqnarray}
  A & = & \frac{1}{R} \times \frac{3\eta}{1-\eta} \times \left(\frac{-A_H^{\text{mat}}}{6\pi q_m\rho_l}\right)^{1/3} ~ \mbox{and} \\
  B & = & \frac{\xi^2}{R^2} \left( \frac{3}{2} \times \frac{\eta}{1-\eta} \times \frac{\mathcal{C}}{2} + 3\left(2-\frac{\pi^2}{8}\right) \times \eta \times \tilde{\rho}_{gb} \right. \nonumber \\  && \left. \hspace{0.8cm} + \frac{1}{3}(4-\pi) \times \tilde{\rho}_{gb}^2 \right) \, ,
\end{eqnarray}
with $\xi = \gamma_{sl}/\rho_sq_m$ and where $c_0 = R_g T_m \rho_0 / q_m \rho_l$ is, up to a constant factor, the density of impurities $\rho_0$ in the fully melted system.

Obviously, Eq.~\eqref{eq_implflNoQAneg} requires that $t > c_0/f_l$ must hold and thus, because $0\le f_l\le 1$, when $t<c_0$ Eq.~\eqref{eq_implflNoQAneg} is not applicable. The case $t<c_0$ corresponds to an undercooling which is smaller than the depression of the bulk freezing point due to the colligative effect originating from the presence of the impurities. Therefore, we must assume a liquid fraction $f_l=1$ when $t<c_0$.

While Eq.~\eqref{eq_implflNoQAneg} does not posses a general analytical solution, it can be solved in the limit of low impurity concentration $c_0$ to yield
\begin{equation}
\label{eq_implflNoQLimc0}
  f_l =  \frac{A}{t^{1/3}} + \frac{B}{t^2} + g_1(\tilde{t}) \times \frac{c_0}{3 t} + g_2(\tilde{t}) \times \frac{c_0^2}{9At^{5/3}} + \mathcal{O}(c_0^3) ,
\end{equation}
with $\tilde{t} = \frac{A}{B}\,t^{5/3}$ where
\begin{eqnarray}
  g_1(\tilde{t}) & = & 1 + \frac{5}{\tilde{t}+1} ~ ~\mbox{and}\\
  g_2(\tilde{t}) & = & 1 + \frac{14 \tilde{t}^2 - 12 \tilde{t} -1 }{(\tilde{t}+1)^3 }.
\end{eqnarray}
Note that the term $\propto c_0$ is sufficient to obtain the correct leading contribution due to impurities in both limiting cases $t \to 0$ and $t \to \infty$. The factor $\propto c_0^2$ becomes a constant in the limit $t\to 0$ where $g_2$ vanishes. The coefficient belonging to $c_0^3$ vanishes as $t^{-7/3}$ in the limit $t\to\infty$ and is thus subdominant to $\frac{B}{t^2}$. 

In the case where dispersion forces are attractive ($A_H^{\text{mat}}>0$) the formation of a premelted layer requires an impurity concentration which is {\em higher} than the one which would be required to prevent the bulk liquid from freezing. Hence, in this case all the liquid water in the matrix pores must originate from liquid veins which form due to curvature induced premelting. In contrast to the premelted layer with $A_H^{\text{mat}}>0$ these veins have formed when $t>c_0/f_l$, 
a temperature for which the premelted layers are still collapsed and do not contribute to the liquid fraction. Thus, premelted layers do not actually form because at higher temperatures, beginning with the veins, the sample completely liquifies. Therefore, the corresponding equation for $f_l$ is 
\begin{equation}
\label{eq_implflNoQApos}
  f_l = \frac{B}{(t-c_0/f_l)^2} \, , 
\end{equation}
which has a simple analytical solution. Again, we examine the leading order contributions in the limiting cases $t\to 0$ and $t\to\infty$ to find
\begin{eqnarray}
  f_l & \stackrel{t\to 0}{\longrightarrow} & \frac{B}{t^2} + \frac{2 c_0}{t} - \frac{c_0^2}{B} + \mathcal{O}(t) ~ ~\mbox{and} \\
  f_l & \stackrel{t\to \infty}{\longrightarrow} & \frac{c_0}{t} + \frac{\sqrt{B c_0}}{t^{3/2}} + \frac{B}{2 t^2} + \mathcal{O}\left(\frac{1}{t^{5/2}}\right) \, , \label{eq_curvOnlytInfty}
\end{eqnarray}
where the latter relation is an extension of Eq.~\eqref{eq_implflNoQLimc0} which is undefined in the limit $t\to\infty$ and $A \to 0$.

\paragraph*{Non-zero surface charge}

\begin{figure}[t]

\begin{center}

\includegraphics[width=.45\textwidth]{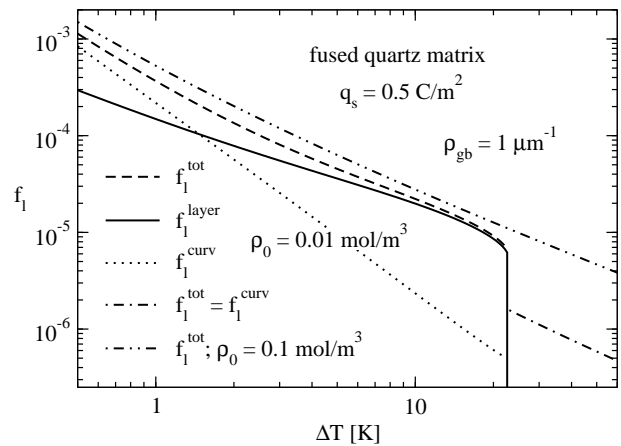}

\caption{Liquid fraction $f_l$ for water in a porous matrix of fused quartz with a grain size $R=500\text{ $\mu$m}$ as a function of the undercooling $\Delta T$. The result for the low impurity density $\rho_0 = 0.01\text{ mol/m$^3$}$ (with respect to the interstitial volume) shows a discontinuity where the quasi-liquid layer between the ice and the matrix collapses. This is accompanied by an increase of the impurity concentration in the remaining (curvature melting induced) liquid which creates a discontinuous increase of $f_l^{\text{curv}}$. For the higher impurity density $\rho_0 = 0.1\text{ mol/m$^3$}$ the discontinuity is not visible in the figure because it occurs at large undercooling ($\Delta T \approx 110\text{ K}$).}

\label{fig_flQuartz}

\end{center}

\end{figure}

In the general case with immobile surface charges ($q_s \neq 0$) we solve Eq.~\eqref{eq_implfl} numerically. In Fig.~\ref{fig_flQuartz} we show results for a matrix 
 of fused quartz with a grain size of $R=500\text{ $\mu$m}$, which may serve as a model for frozen soils. The surface charge density is chosen to be $q_s = 0.5\text{ C/m$^2$}$ which has the correct order of magnitude for mineral/water interfaces at high pH \cite{DoCr05}. The ice which fills the voids of the porous matrix is taken to be polycrystalline with a grain boundary density $\rho_{\text{gb}} = 1\text{ $\mu$m$^{-1}$}$. A comparison of Eqs.~\eqref{eq_flmatlayer} and \eqref{eq_gblayer} shows that for the given $\rho_{\text{gb}}$ the surface area provided by grain boundaries is by a factor $100$ larger than the surface area of the porous matrix. Therefore, we expect the impurity concentration in the sample to be very low when grain boundaries are melted and, according to Eq.~\eqref{eq_dminTmaxNiInfty}, the maximal undercooling $\Delta T_{\text{max}}$ for which quasi-liquid interfacial water at the grain boundaries is stable should be low. An upper limit for $\Delta T_{\text{max}}$ can be calculated by assuming that all the impurities in the sample are located at the premelted grain boundaries. 
For the values of $\rho_0$ from Fig.~\ref{fig_flQuartz} we use  $N_i = \rho_0/\rho_{\text{gb}}$ to obtain a range $N_i = 0.01-0.1\text{ $\mu$mol/m$^2$}$ and, with the aid of  Eq.~\eqref{eq_dminTmaxNiInfty} this corresponds to a range in 
 $\Delta T_{\text{max}}$ of $0.01-0.26\text{ K}$.  The range of $\Delta T_{\text{max}}$ uses the Hamaker constant $A_H = 3.3\times 10^{-22}\text{ J}$ computed from complete dispersion force theory \cite{DzLiPi61} in the limit of thin layers with the data for the dielectric functions from Elbaum and Schick \cite{ElbSch91}. Including a typical surface charge density of $q_s = 0.01\text{ C/m$^2$}$ for the grain boundary \cite{BeWe04} lowers these values to $\Delta T_{\text{max}} = 0.01-0.20\text{ K}$.  Hence, the grain boundaries are always collapsed to zero film thickness for the range of $\Delta T$ considered in the figure. A prominent feature of  $f_l = f_l^{\text{tot}}$   with $\rho_0 = 0.01\text{ mol/m$^3$}$ in the fused quartz matrix is the discontinuity at $\Delta T = 22.6\text{ K}$ where $f_l$ drops by almost an order of magnitude. This is the signature of the collapse of the premelted layer at the ice/fused quartz interface described in detail in Section~\ref{sec_resLayer}. In the strongly undercooled, layer  collapsed system, the only contribution to $f_l$ comes from curvature induced melting;  $f_l^{\text{curv}} = f_l^{\text{contact}} + f_l^{\text{gbmat}} + f_l^{\text{veins}}$ (cf.\ Eq.~\ref{eq_fltot}), which discontinuously increases at the undercooling where the premelted layer collapses. This is because, due to the efficient rejection of impurities by the ice lattice, the impurity concentration in the remaining liquid increases discontinuously when the liquid layers collapse. 
At low undercooling curvature melting becomes increasingly important exceeding the contribution from interfacial premelting $f_l^{\text{layer}} = f_l^{\text{mat}}$.  However, at larger undercooling, close to the location of layer collapse, curvature induced melting yields only a minute contribution to $f_l$. The importance of curvature induced melting at low undercooling is reflected in Eq.~\eqref{eq_implflNoQLimc0} where the curvature term $B/t^2$ dominates for $t \to 0$. In the opposite limit of large undercooling the analytical result Eq.~\eqref{eq_curvOnlytInfty} applies. The leading term $c_0/t$ originates from the colligative effect and   curvature induced melting enters only through the subdominant term. Hence, for large undercooling the principal contribution to the liquid fraction is the colligative effect. However,  liquid water is located in  regions of high curvature; contact of matrix spheres, contact lines of grain boundaries and matrix spheres and liquid veins. Figure~\ref{fig_flQuartz} also shows $f_l$ in the same system with a higher impurity concentration ($\rho_0=0.1\text{ mol/m$^3$}$). Here the collapse of the liquid layer is moved to a lower temperature ($\Delta T \approx 110\text{ K}$) so that no discontinuity is observed in the range of undercooling considered in the figure.

\begin{figure}[t]

\begin{center}

\includegraphics[width=.45\textwidth]{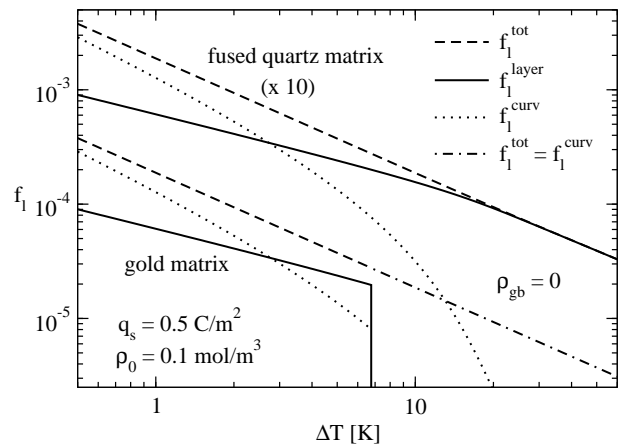}

\caption{Liquid fraction $f_l$ for the gold matrix compared to the fused quartz matrix as a function of the undercooling $\Delta T$ for $\rho_0 = 0.1\text{ mol/m$^3$}$. Importantly, although the Hamaker constants for the two materials differ by almost two orders of magnitude, which results in a layer collapse for the gold matrix at $\Delta T \approx 7\text{ K}$, the curves for the total liquid fraction are almost identical.  Note that the curves for fused quartz have been shifted for clarity. In particular, the discontinuity of $f_l^{\text{tot}}$ at the location of layer collapse is too small to be resolved in the figure, which results from the high impurity concentration, see Eq.~\eqref{eq_deltafl}. The strength of dispersion forces is important only for thin premelted layers, i.e.\ in the immediate vicinity of layer collapse. The discontinuity for fused quartz is not visible in the figure because it occurs at large undercooling ($\Delta T \approx 103\text{ K}$).}

\label{fig_flGoldvsQuartz}

\end{center}

\end{figure}

In Fig.~\ref{fig_flGoldvsQuartz} many of the same conditions as in Fig.~\ref{fig_flQuartz} are considered ($R = 500\text{ $\mu$m}$, $q_s = 0.5\text{ C/m$^2$}$, $\rho_0=0.1\text{ mol/m$^3$}$) the principal difference being that the pore ice is single-crystalline ($\rho_{\text{gb}} = 0$). We find that $f_l^{\text{tot}}$ in the fused quartz matrix is qualitatively the same as for the polycrystalline ice (see Fig.~\ref{fig_flQuartz}) in that it does not show liquid layer collapse on the scale of the plot for $\rho_0=0.1\text{ mol/m$^3$}$. When replacing the matrix with gold, which has stronger attractive dispersion forces at the ice-gold interface, we see a collapse of the premelted layer at $\Delta T \approx 7\text{ K}$. This is accompanied by, as in the case of layer collapse for the fused quartz matrix at $\rho_0=0.01\text{ mol/m$^3$}$ (Fig.~\ref{fig_flQuartz}), a discontinuous increase of the liquid fraction $f_l^{\text{curv}}$. As for the fused quartz matrix, for sufficiently low undercooling that the premelted layer has collapsed, the only contribution to $f_l^{\text{tot}}$ comes from curvature induced melting; $f_l^{\text{tot}} = f_l^{\text{curv}}$. While, on the scale of the plot $f_l^{\text{tot}}$ does not display a discontinuity at the point of layer collapse, there is a discontinuity $\Delta f_l$ which can be seen by expanding the full result in the limit of large impurity concentration $c_0$ 
\begin{equation}
\label{eq_deltafl}
  \Delta f_l = \frac{1}{2} \times f_l^{\text{layer}} \times \left(\frac{B}{c_0 t_{\text{coll}}}\right)^{1/2} + \mathcal{O}\left(\frac{1}{c_0^{3/2}}\right) \, ,
\end{equation}
where $t_{\text{coll}}$ is the undercooling at layer collapse and $f_l^{\text{layer}}$ the liquid fraction from interfacial melting at a temperature slightly above collapse $t = t_{\text{coll}} - \delta t$, with $\delta t > 0$.  Clearly, $\Delta f_l \to 0$ for $c_0 \to \infty$.  

An important result seen in Fig.~\ref{fig_flGoldvsQuartz} is that although the Hamaker constants for gold and fused quartz differ by almost two orders of magnitude, the results for $f_l$ are almost identical over a large range of $\Delta T$; the deviation increases from $0.02\text{ \%}$ to $0.4\text{ \%}$ at layer collapse and is $6\text{ \%}$ for $\Delta T = 60\text{ K}$. This is because dispersion forces are only relevant for thin premelted layers and hence at temperatures slightly above layer collapse. However, when layer collapse occurs, according to Eq.~\eqref{eq_deltafl} the high impurity concentration, combined with single-crystallinity (i.e.\ small $B$), insures that the Debye length and the discontinuity of $f_l$ are small.  Therefore, the deviation between the results for the two different matrix materials remains small. The conclusions drawn from Fig.~\ref{fig_flGoldvsQuartz} are important to note for experimental determination of $f_l$ in different materials because they imply that even though the underlying microscopic behavior of two samples may fundamentally differ (interfacial premelting vs.\ collapsed premelted layer) this may not  be resolved in the integrated/homogenized quantity $f_l$. Spatially resolved measurements may be required to detect characteristic transitions such as the collapse of interfacially premelted layers.

\paragraph*{Effect of polycristallinity}

\begin{figure}[t]

\begin{center}

\includegraphics[width=.45\textwidth]{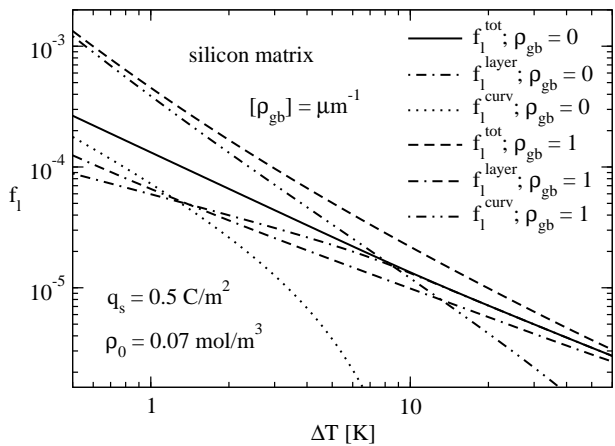}

\caption{Liquid fraction $f_l$ for a matrix consisting of silicon spheres with radius $R = 500\text{ $\mu$m}$ as a function of the undercooling $\Delta T$. Two different cases concerning the ice, which is contained in the matrix, are considered: polycrystalline ice ($\rho_{\text{gb}} = 1\text{ $\mu$m$^{-1}$}$) and single-crystalline ice ($\rho_{\text{gb}} = 0$). The polycrystalline sample always has a higher liquid fraction, the difference with the single-crystalline sample being most pronounced at small undercooling. In this regime $f_l$ in the polycrystalline sample is almost entirely made up of $f_{\text{curv}}$, which mainly contains contributions from liquid veins in the ice. An interesting consequence of the non-monotonic dependence of the premelted layer thickness on the impurity density $N_i$ can be observed: although $N_i$ is always larger in the single-crystalline sample than in the polycrystalline sample, the curves for $f_l^{\text{layer}}$ (which are proportional to the layer thicknesses) intersect.}

\label{fig_flSilicon}

\end{center}

\end{figure}

Finally, we study the influence of the degree of polycrystallinity of the interstitial ice for the case where the porous matrix consists of silicon, which leads to dispersion forces that are repulsive ($A_H<0$) and prevent the interfacially premelted layer from collapsing. As in the previous examples the grain size of the matrix is $R=500\text{ $\mu$m}$  and as in Fig.~\ref{fig_flQuartz} the grain boundaries are collapsed.  Thus, Fig.~\ref{fig_flSilicon} shows the curves for $f_l=f_l^{\text{tot}}$,  the interfacial melting ($f_l^{\text{layer}}$) and curvature induced melting ($f_l^{\text{curv}}$)  contributions for single-crystalline ($\rho_{\text{gb}}=0$) and polycrystalline ice ($\rho_{\text{gb}}=1\text{ $\mu$m$^{-1}$}$). 
As expected, the polycrystallinity leads to a higher liquid fraction $f_l$.  In particular, for small undercooling $\Delta T$, where curvature induced melting is enhanced and hence liquid veins greatly contribute, $f_l$ consists almost entirely of $f_l^{\text{curv}}$. The values of $f_l^{\text{curv}}$ for the single crystal  ice, for which contributions derive only from the contact points between matrix spheres, are always by at least an order of magnitude lower than for the polycrystalline ice case. This is not as obvious as it may first appear. Due to the higher liquid fraction in polycrystalline ice the impurity concentration $\rho_i$ is smaller than for the single crystal case. Thus,  polycrystalline ice has a smaller ice-water radius of curvature (see Eq.~\ref{eq_rcurvmelt}). Consequently, the volume of liquid contained for example at the contact point between two matrix spheres, is smaller for the polycrystalline case than for the single crystal case. It is the 
grain boundary density $\rho_{\text{gb}}$, which causes $B$ to be larger for the polycrystalline ice, that insures $f_l^{\text{curv}}$ will be also larger, as indicated by Eq.~\eqref{eq_implflNoQLimc0}. In addition to the cases studied in Section~\ref{sec_resLayer}, another interesting manifestation of the non-monotonic dependence of the premelted layer thickness on the impurity concentration for the systems with interfacial charges can be seen in Fig.~\ref{fig_flSilicon}. Although the impurity concentration $\rho_i$ and the impurity density $N_i$ are always larger in the single-crystal case, the liquid fractions $f_l^{\text{layer}}$, which from Eq.~\eqref{eq_flmatlayer} are proportional to the corresponding layer thicknesses, intersect at an undercooling $\Delta T_{\text{int}} = 1.38\text{ K}$. Thus for $\Delta T < \Delta T_{\text{int}}$ the larger impurity density of a single crystal  sample leads to a smaller layer thickness than the polycrystalline case. It is only for $\Delta T > \Delta T_{\text{int}}$ that the usual increase of the layer thickness with increasing impurity concentration is observed. This complex behavior provides further evidence for the observation (Fig.~\ref{fig_flGoldvsQuartz}), that measuring the integrated/homogenized quantity $f_l$ alone does not yield satisfactory information about the microscopic structure and behavior of the sample, in particular the rich and interesting temperature dependence of the thickness of the premelted layer in a frozen matrix containing ice with different degrees of polycrystallinity.  By parity of reasoning, inclusion of just one or two microscopic phenomena in a theory that is intended to predict the effective medium behavior of a porous medium, will provide irrelevant and misleading predictions unless the proper asymptotic limits of a complete theory are assured. 

\section{Summary and conclusion}

\label{sec_sumconc}

In this work we construct a theory for the premelting of ice in a porous matrix (Section~\ref{sec_theory}), which is modeled as a random close packing of monodisperse spheres. The theory includes premelting at the matrix-ice interface, grain boundary premelting, as well as curvature and impurity induced premelting. The latter includes premelting in liquid veins, at contact lines of the matrix with a grain boundary, and at the contact points between two matrix spheres. The matrix is characterized by the sphere radius $R$ and the matrix material, which enters the theory via the Hamaker constant $A_H$ for the ice/liquid/matrix layered system. In addition to dispersion forces at the ice-matrix interface, which, depending on the matrix material, either stabilize or destabilize the premelted quasi-liquid layer, we include the effect of electrostatic forces at the interface due to adsorbed surface charges at the interface along the lines of our previous work Ref.~\onlinecite{We99}. The theory includes the effect of charged impurities in the premelted water both on the thickness of the premelted layers (grain boundaries, ice-matrix interface) as well as on the curvature induced melting. The main idea of the present approach is to make use of the small segregation coefficient of ice which leads to an enrichment of impurities in the premelted liquid (making the impurity concentration a function of the fraction of premelted liquid) and of the fact that regions of premelted liquid in a sample are all interconnected and therefore the impurity concentration is homogeneous. This approach allowed us to derive an implicit equation for the liquid fraction $f_l$ as a function of the undercooling $\Delta T$, Eq.~\eqref{eq_fltot}, which we analyze for general parameters by considering limiting cases where analytical solutions are possible (Section~\ref{sec_analysis}). To understand each contribution we began by performing a careful study of interfacial melting, analyzing an implicit equation for the thickness $d$ of the premelted layer as a function of $\Delta T$, Eq.~\eqref{eq_explinterf}, for general parameters. In particular, we addressed the discontinuous collapse of the premelted layer under certain conditions and the non-monotonic dependence of $d$ on the impurity concentration (cf.\ Ref.~\onlinecite{We99}). However, in the general case both Eq.~\eqref{eq_explinterf} and Eq.~\eqref{eq_fltot}, the latter of which makes use of the former, have to be solved simultaneously through a numerical scheme. In so doing we chose physically relevant parameters of interest to a wide community of scientists.  For the porous matrix  we use $R=500\text{ $\mu$m}$ and considered two materials (fused quartz and gold) that yield attractive dispersion forces at the ice/matrix interface, thereby inhibiting the formation of a premelted quasi-liquid layer, and one material (silicon) which enhances the premelted layer. The surface charge at the interface is taken to be $q_s = 0.5\text{ C/m$^2$}$. The impurity concentrations lie in the range of $0.01-0.1\text{ mol/m$^3$}$ defined with respect to the interstitial volume, and we consider both single- and polycrystalline ice. We calculate $f_l$ for $\Delta T$ in the range of $0.5-60\text{ K}$ and obtain liquid fractions which range   between $10^{-6}$ and $10^{-3}$ of the interstitial volume. 

For the parameters under consideration the grain boundaries are always of zero thickness (i.e., collapsed), and hence only the collapse of the quasi-liquid layer at the ice-matrix interface is observed in $f_l$ (see Fig~\ref{fig_flQuartz} for the case of the fused quartz matrix). When the temperature of the sample is decreased below the point of layer collapse, only curvature induced premelting contributes to $f_l$. While dispersion forces are too weak in the fused quartz system to significantly influence the results, in the case of the gold matrix dispersion forces are sufficiently strong   to drive layer collapse, as opposed to the surface charge triggered layer collapse in the fused quartz system (see Fig.~\ref{fig_flGoldvsQuartz} for a comparison of the two cases). An important conclusion from Fig.~\ref{fig_flGoldvsQuartz} is that for high impurity concentration and/or low degree of polycrystallinity the characteristic discontinuity of $f_l$ upon layer collapse is minute and most likely not resolved in experiments.  Thus, transitions that would be dramatic and easily detected  at planar interfaces will be much more challenging to extract from  measurements of the integrated/homogenized quantity $f_l$. Finally, we study the influence of the degree of polycrystallinity on $f_l$ for the case of the silicon matrix where, due to the repulsive dispersion forces, layer collapse does not occur  (Fig.~\ref{fig_flSilicon}). The results reveal an interesting consequence of the non-monotonic dependence of $d$ on the impurity concentration. Although the actual impurity concentration is higher in the single-crystalline sample than in the polycrystalline sample at all temperatures, the thickness of the premelted layer can be both larger and smaller in either case, depending on the temperature range. This peculiarity is obscured by the curvature induced contribution to $f_l$ providing further evidence that additional measurements are required $f_l$   to capture the physics controlling the sample microstructure.

We view our approach as having provided a thorough theory of premelting in a porous medium which can be applied to a plethora of experimental situations and material systems. For the sake of brevity and clarity, we   presented examples of specific materials systems; gold, fused quartz, and silicon.  However, we chose these materials because their dielectric properties, as reflected  in their Hamaker constants, span all the range of materials as diverse as sapphire, polystyrene, and polyvinylchloride \cite{WiWeElSch95}.   We did not explore the effect of varying the matrix sphere radius $R$, its roughness or its  surface charge density $q_s$. The value $R=500\text{ $\mu$m}$ was chosen with the example of frozen soils in mind, while $q_s=0.5\text{ C/m$^2$}$ provides a realistic order of magnitude as inferred from experimental data for mineral-water interfaces at large pH \cite{DoCr05}. However, not much is known experimentally about $q_s$ in the ice/liquid/matrix systems and the simplifying assumptions of symmetry (same charge at the ice-liquid and liquid-matrix interfaces) as well as independence of $q_s$ on the impurity concentration and layer thickness are not yet established experimentally on planar systems. Nevertheless, given the importance of the impurity concentration dependent screening length for regulating the strength and range of the electrostatic force, the precise value of $q_s$ may be considered secondary.

An important question for future research concerns the dynamical evolution of the porous matrix system in response to changes in the control parameters, in particular the sample undercooling $\Delta T$. In order to guarantee a homogeneous impurity concentration throughout a sample, $\Delta T$ must be changed on a time scale that is slow compared to the typical impurity diffusion time. An even more ambitious question concerns impurity diffusion during layer collapse. The situation where the premelted quasi-liquid layer between the ice and the matrix collapses, which forces the impurities to migrate to remaining liquid regions with curvature induced melting constitutes a challenging task for further investigations.  

\begin{acknowledgements}

The authors have benefitted from discussions with J.G. Dash, S.S.L. Peppin, M. Spannuth, and E. Thomson at various stages of this work. We acknowledge the U.S. National Science Foundation Grant No. OPP0440841, the Department of Energy Grant No. DE-FG02-05ER15741, the Helmholtz Gemeinschaft Alliance ``Planetary Evolution and Life'', and Yale University for generous support of this research.  

\end{acknowledgements}

\appendix

\section{Melting volume at a grain boundary--matrix contact line }

\label{app_grainbmat}

\begin{figure}[t]

\begin{center}

\includegraphics[width=.45\textwidth]{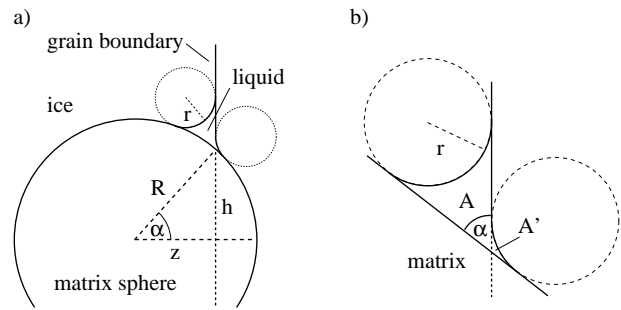}

\caption{a) Geometry of a matrix sphere intersecting with a grain boundary which is located at distance $z$ from the sphere center. b) Zoom on the curvature melting induced liquid ``vein''. As $r\ll R$ the matrix surface can be considered as flat in the vicinity of the liquid vein.}

\label{fig_grainbmat}

\end{center}

\end{figure}

We consider a matrix sphere that is embedded at the interface between two ice grains. The grains are assumed to have a planar grain boundary and to be much larger than the matrix sphere. Let the distance between the sphere center and the grain boundary be $z<R$. Curvature induced melting leads to a circular ``vein'' along the line where the grain boundary and the sphere intersect (Fig.~\ref{fig_grainbmat}a). Because $r$ is assumed to be much smaller than $R$ we can consider the matrix surface as being planar in the vicinity of the vein as shown in Fig.~\ref{fig_grainbmat}b.
The angle between the grain boundary and the matrix sphere is
\begin{equation}
  \alpha = \arccos\left( \frac{z}{R} \right) \, .  
\end{equation}
Using elementary geometry we obtain the surface areas
\begin{eqnarray}
  A & = & \cot\left(\frac{\alpha}{2}\right) r^2 - \left(\frac{\pi}{2}-\frac{\alpha}{2}\right) r^2 \, , \\
 A' & = & \tan \left(\frac{\alpha}{2}\right) r^2 - \frac{\alpha}{2} \, r^2
\end{eqnarray}
and the total surface area
\begin{equation}
  A_{\text{tot}} = A + A' = \left(\frac{2}{\sin\alpha}-\frac{\pi}{2}\right) r^2 \, .
\end{equation}
Accordingly, the volume of premelted water for the configuration in Fig.~\ref{fig_grainbmat} is $2\pi h A_{\text{tot}}$. Taking all possible distances of the matrix sphere to the grain boundary into account yields an average volume $\bar{v}$ of premelted liquid as
\begin{eqnarray}
  \bar{v} & = & \frac{1}{R} \int_0^R dz 2\pi h A_{\text{tot}} \nonumber \\
            & = &  \frac{1}{R} \int_0^R dz 2\pi R \sin\alpha\left(\frac{2}{\sin\alpha}-\frac{\pi}{2}\right) r^2   \nonumber \\
            & = & 2\pi r^2 \int_0^R dz \left( 2 - \frac{\pi}{2} \sqrt{1-\frac{z^2}{R^2} } \right) \nonumber \\
            & = & 2\pi r^2 R \left( 2 - \frac{\pi^2}{8}\right) \, .
\end{eqnarray}

Now consider a random close packing with packing fraction $\eta$ (the porous medium), which contains ice with two different crystallographic orientations being separated by a planar grain boundary. The volume fraction $\eta$ is also the surface fraction of the grain boundary which is cut out in the form of circles with radii between $0$ and $R$ due to the presence of the matrix spheres. Given that the average surface area per intersecting sphere is
\begin{equation}
  \bar{a} = \frac{1}{R} \int_0^R dz \pi (R^2-z^2) = \frac{2}{3}\pi R^2
\end{equation}
a sample of volume $V_s = A_s x_s$ with a grain boundary perpendicular to the $x$-axis contains $\eta A_s/\bar{a}$ intersection circles of matrix spheres with the grain boundary, if $A_s$ is sufficiently large. If the sample contains the total grain boundary surface $\rho_{\text{gb}}' V_s$ this corresponds to a net surface $A_s= \rho_{\text{gb}}' V_s/(1-\eta)$ and hence a density of intersection circles
\begin{equation}
  n_c = \frac{ \eta \rho_{\text{gb}}'}{(1-\eta)\bar{a}} = \frac{ \eta \rho_{\text{gb}}}{\bar{a}} \, ,
\end{equation}
where $\rho_{\text{gb}}$ is the grain boundary density with respect to the interstitial volume. The volume of premelted liquid along the intersection circles (per unit volume of the sample) is $\bar{v} n_c$. Or, relative to the interstitial volume:
\begin{equation}
  \frac{\bar{v} n_c}{1-\eta} = 3 \left( 2 - \frac{\pi^2}{8} \right) \times \frac{\eta}{1-\eta} \times \tilde{\rho}_{\text{gb}} \times \left(\frac{r}{R} \right)^2 \, .
\end{equation}

\end{document}